\documentclass[aps,floatfix,twocolumn,superscriptaddress]{revtex4}
\usepackage
{graphicx,subfigure,graphics,epsfig,amsmath,amssymb}
\usepackage{graphicx}
\usepackage{graphics,epsfig}
\usepackage{amssymb}
\usepackage{symb}

\begin{document}
\title{Is the Dynamics of Scaling Dark Energy Detectable?}

\author{Bruce A. Bassett}
\affiliation{South African Astronomical Observatory, Observatory, Cape Town, South Africa}
\affiliation{Department of Mathematics and Applied Mathematics, University of Cape Town, Rondebosch, 7700, Cape
Town, South Africa}

\author{Mike Brownstone}
\affiliation{National Astrophysics and Space Science Programme}  \affiliation{Department of Mathematics and
Applied Mathematics, University of Cape Town, Rondebosch, 7700, Cape Town, South Africa}

\author{Antonio Cardoso}
\affiliation{Institute of Cosmology \& Gravitation, University of Portsmouth, Portsmouth~PO1~2EG, UK}

\author{Marina Cort\^{e}s}
\affiliation{Astronomy Centre, University of Sussex, Brighton BN1 9QH, United Kingdom}

\author{Yabebal Fantaye}
\affiliation{National Astrophysics and Space Science Programme}  \affiliation{Department of Mathematics and
Applied Mathematics, University of Cape Town, Rondebosch, 7700, Cape Town, South Africa}

\author{Ren\'{e}e Hlozek}
 \affiliation{Department of Mathematics and Applied
Mathematics, University of Cape Town, Rondebosch, 7700, Cape Town, South Africa} \affiliation{South African
Astronomical Observatory, Observatory, Cape Town, South Africa}

\author{Jacques Kotze}
\affiliation{Department of Mathematics and Applied Mathematics, University of Cape Town, Rondebosch, 7700, Cape
Town, South Africa}

\author{Patrice Okouma}
\affiliation{National Astrophysics and Space Science Programme}  \affiliation{Department of Mathematics and
Applied Mathematics, University of Cape Town, Rondebosch, 7700, Cape Town, South Africa}

\begin{abstract}
We highlight the unexpected impact of nucleosynthesis and other early universe constraints on the detectability
of scaling quintessence dynamics at late times, showing that such dynamics may well be invisible until the
unveiling of the Stage-IV dark energy experiments (DUNE, JDEM, LSST, SKA). Nucleosynthesis strongly limits
potential deviations from $\Lambda$CDM. Surprisingly, the standard Chevallier-Polarski-Linder (CPL) parametrisation, $w(z) = w_0 + w_a
z/(1+z)$, cannot match the nucleosynthesis bound for minimally coupled scaling fields. Given that such models
are arguably the best-motivated alternatives to a cosmological constant these results may significantly impact
future cosmological survey design and imply that dark energy may well be dynamical even if we do not detect any
dynamics in the next decade.
\end{abstract}

\maketitle
\section{Introduction}

The quintessential enigma of modern cosmology is that of dark energy - the shadowy source of the accelerated
expansion of the cosmos. Einstein's ``greatest blunder" -- the cosmological constant, $\Lambda$, remains the
most economical explanation for this observed acceleration, but more exciting alternatives could imply the
existence of new types of matter, modifications of the Einstein equations or even violation of the Copernican
principle.

Of these however, one can argue that only scalar fields are consistent with current constraints while
simultaneously being theoretically well-founded. The more exotic alternatives either are claimed to have
theoretical pathologies or are likely to be indistinguishable from the vanilla $\Lambda$CDM model which, given
all current data, remains flavour of the month \cite{wmap3}. Models in the former category include K-essence
\cite{kessence1} which may exhibit an unbounded adiabatic speed of sound \cite{k-essence_sound} and the DGP
model \cite{dgp1}, which has some theoretical problems with ghosts (see \cite{dgp_ghosts} and references
therein) while those in the latter class include $f(R)$ modifications of gravity \cite{SPH} and unified dark
energy \cite{sand}. Of the scalar fields, arguably the best-motivated and most compelling are the scaling
quintessence models \cite{ratra,copeland_tsujikawa, ferreira97}. We consider such scaling models with a
redshift-dependent equation of state parameter $w(z) = p/\rho$ that tracks the dominant energy density component
of the cosmos ($w=\frac{1}{3},0$) until a redshift $z = z_t$ at which point it undergoes a transition to $w < 0$
which triggers the onset of acceleration.

Big Bang Nucleosynthesis (BBN) provides strong constraints on the energy density of dark energy during the
radiation dominated era at a temperature of $T \sim 1 {\rm MeV}$, implying that $ \Omega_{\rm DE}(T \sim 1 {\rm
MeV}) < \epsilon = 0.045 $ at $2\sigma$ \cite{Bean:2001wt}. If the scalar field has reached the scaling
attractor solution at the time of BBN, this early universe constraint on the energy density of the dark energy
will be carried forward until the field exists the scaling regime (at $z_t$). The time at which the attractor
for the scalar field is approached depends on the details of the physics before the radiation dominated epoch.
But in \cite{ferreira97} it was shown that, for a typical inflationary model with the usual method of reheating,
the field will approach the attractor long before nucleosynthesis. Therefore we assume that the field is already
scaling with the background fluid at the time of BBN. In fact, our results are unchanged even if scaling only
occurs by decoupling, since similar or better constraints on the dark energy density exist from the Cosmic
Microwave Background (CMB), $\Omega_{\rm DE}(T\sim 1 {\rm eV})< 0.04$ \cite{doran_cmb1}. Constraints on $w(z)$
from BBN taken together with recent data require that $z_t > 5$ \cite{longde}.

\begin{figure}[htbp!]
\begin{center}
\epsfxsize=3.4in \epsffile{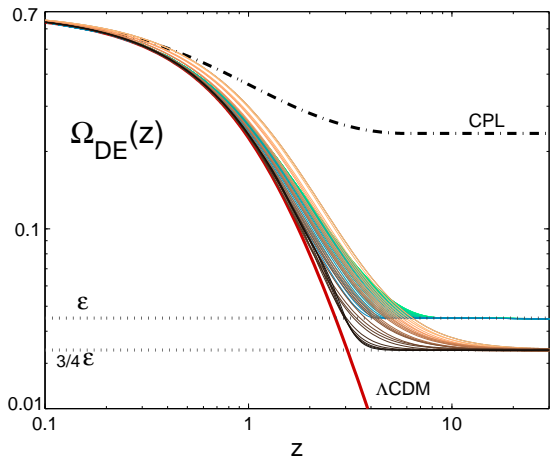} \caption{{\bf Evolution of ${\mathbf\Omega_{\rm \bf DE}\mathbf{(z)}}$} for the
models we consider, showing their approach to the BBN limits of $\epsilon = \Omega_{\rm DE}(z=z_{\rm{\tiny
BBN}}) = 0.045$ for the polynomial parameterisation of $w(z)$ and $\frac{3}{4}\epsilon$ for the double
exponential potential. For comparison we also show the curves for $\Lambda$ and the CPL \cite{cp} $w(z)$ with the lowest
asymptotic value of $\Omega_{\rm DE}$ in this model while still assuming $w\geq -1$, showing its inability to
match the BBN constraint and describe a canonical scalar field. Figure ~(\ref{group1}) shows the corresponding
observational quantities for the scaling quintessence models. \label{ode}}
\end{center}
\end{figure}


\begin{figure*}[t]
\begin{flushleft}
$\begin{array}{@{\hspace{-0.05in}}l@{\hspace{-1in}}l@{\hspace{-0.2in}}l} \multicolumn{1}{l}{\mbox{\bf \large{~~~~~Polynomial
w(z)}}}& &  \\ [0.0cm] \epsfxsize=2.5in \epsffile{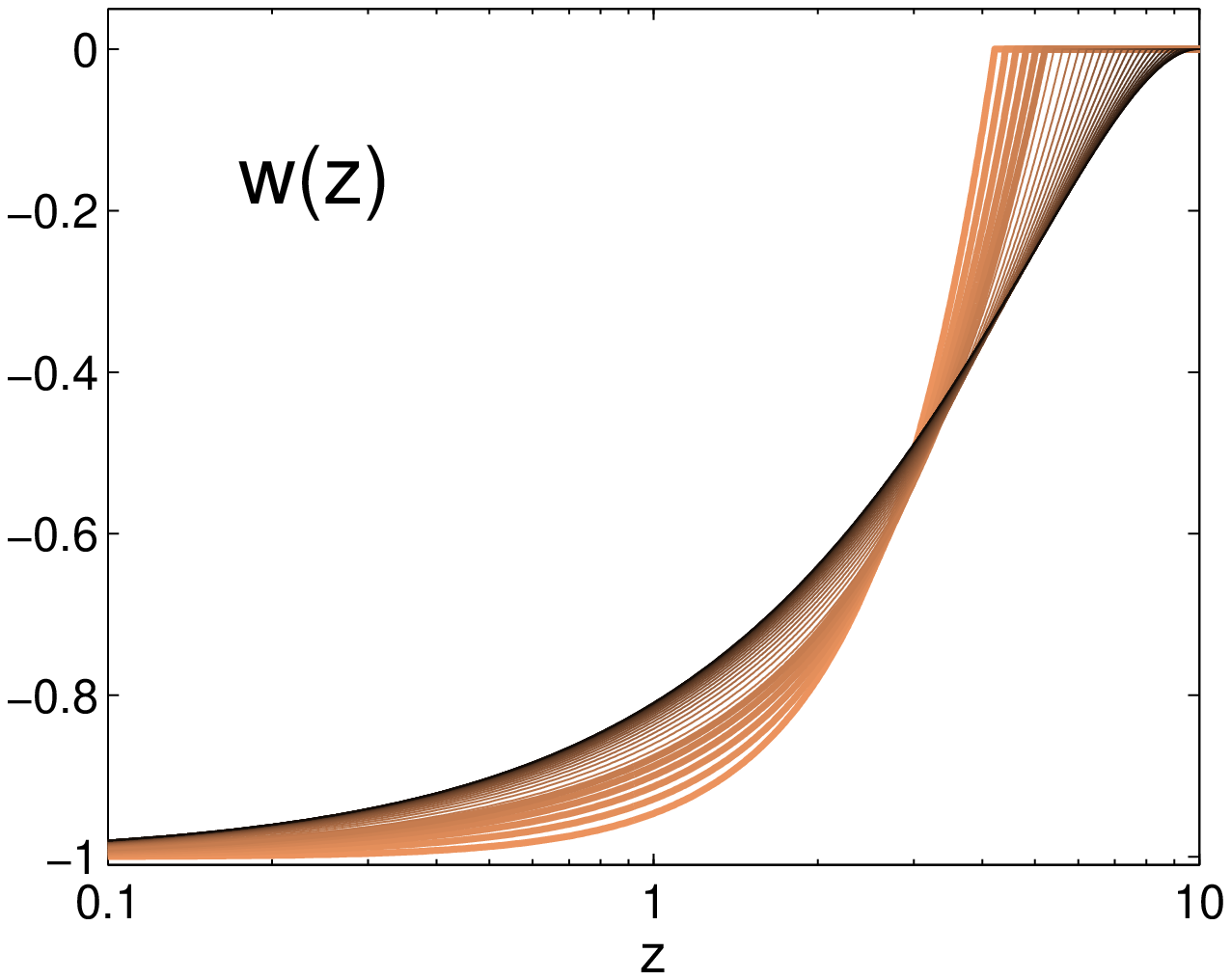} &
    \epsfxsize=2.5in
    \epsffile{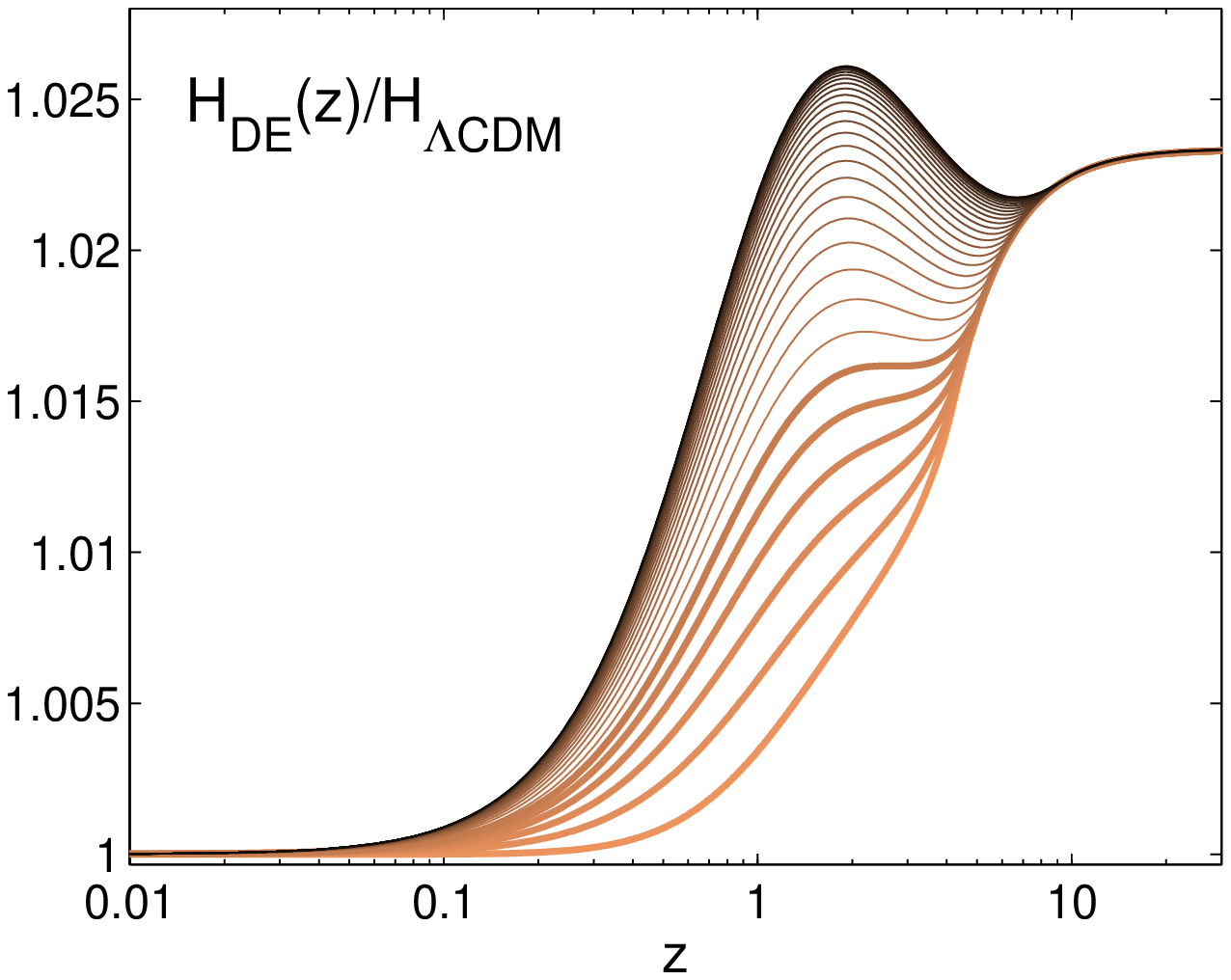}&
        \epsfxsize=2.5in
        \epsffile{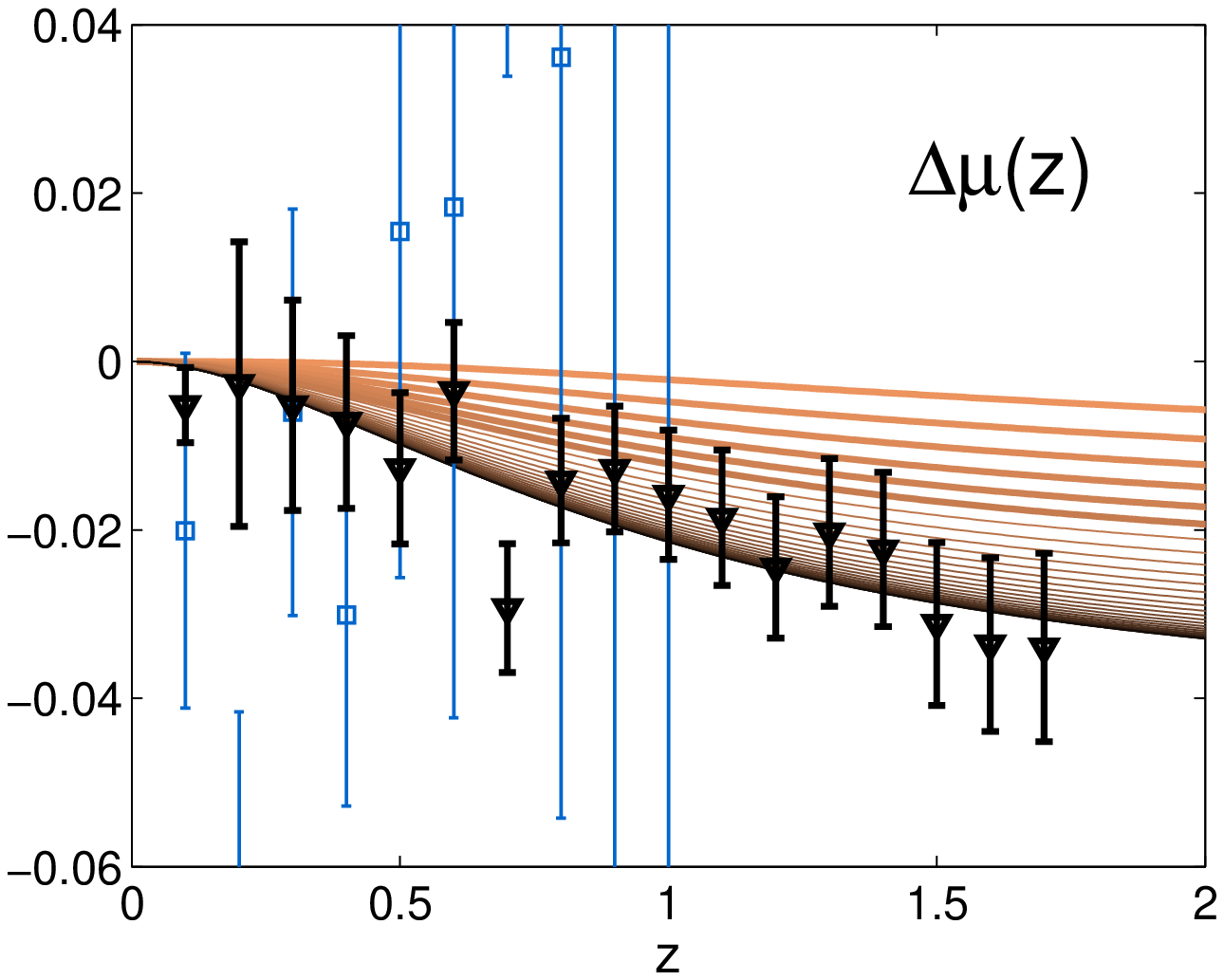}  \\ [0.0cm]

    \multicolumn{1}{l}{\mbox{\bf \large{~~~~~Double Exponential~~~~~~}}}& & \\ [0.0cm]
\epsfxsize=2.5in \epsffile{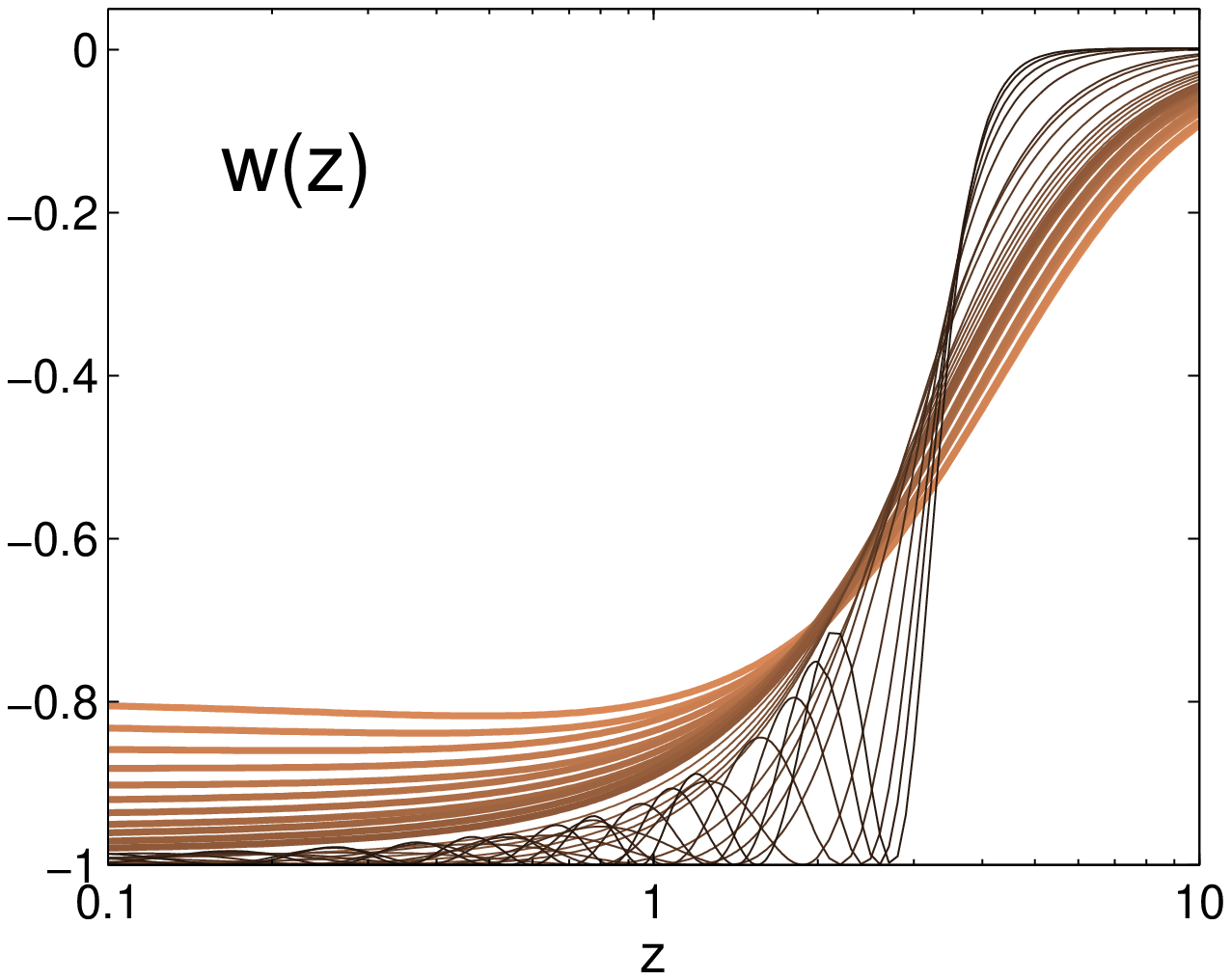}&
    \epsfxsize=2.5in
    \epsffile{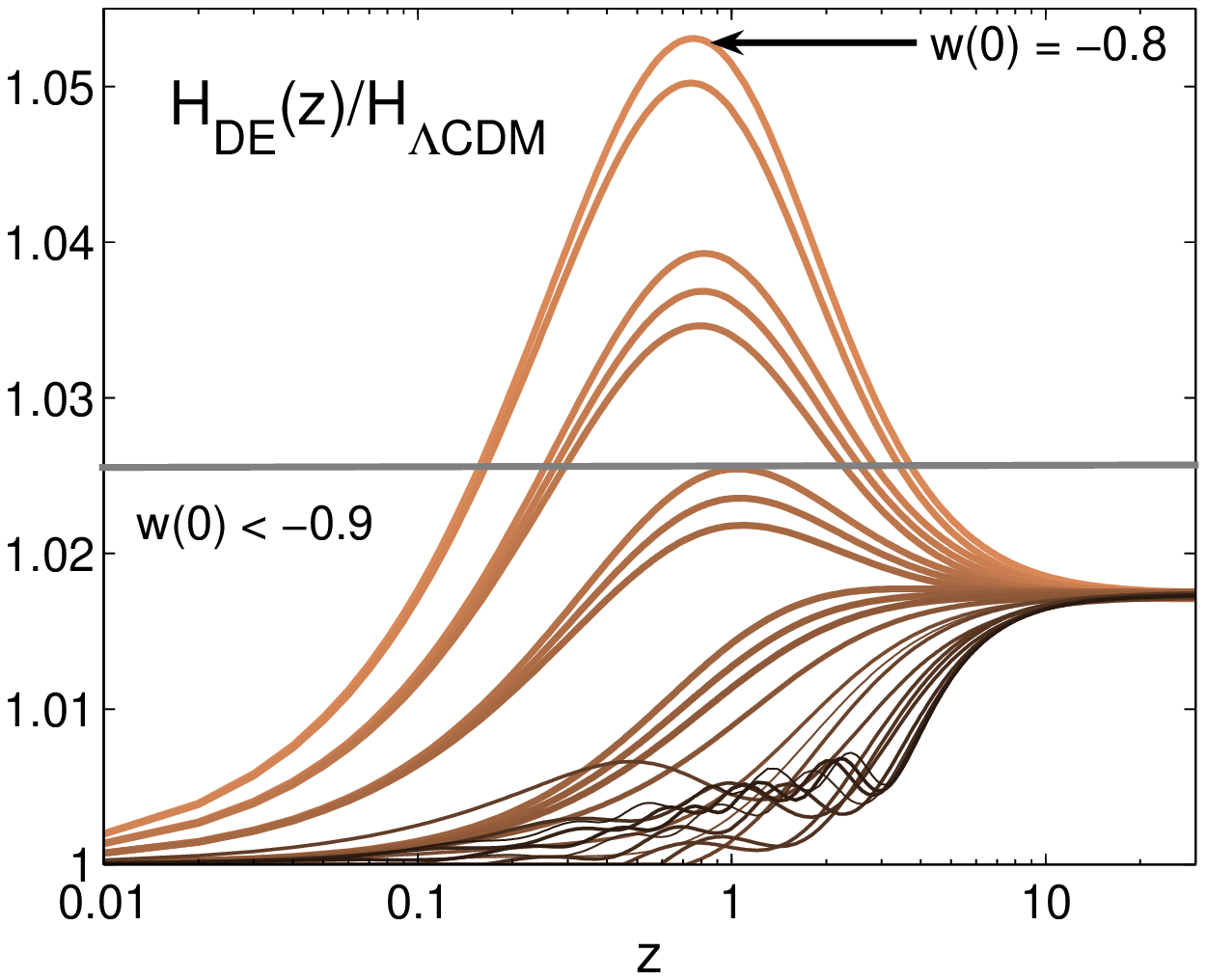} &
        \epsfxsize=2.5in
        \epsffile{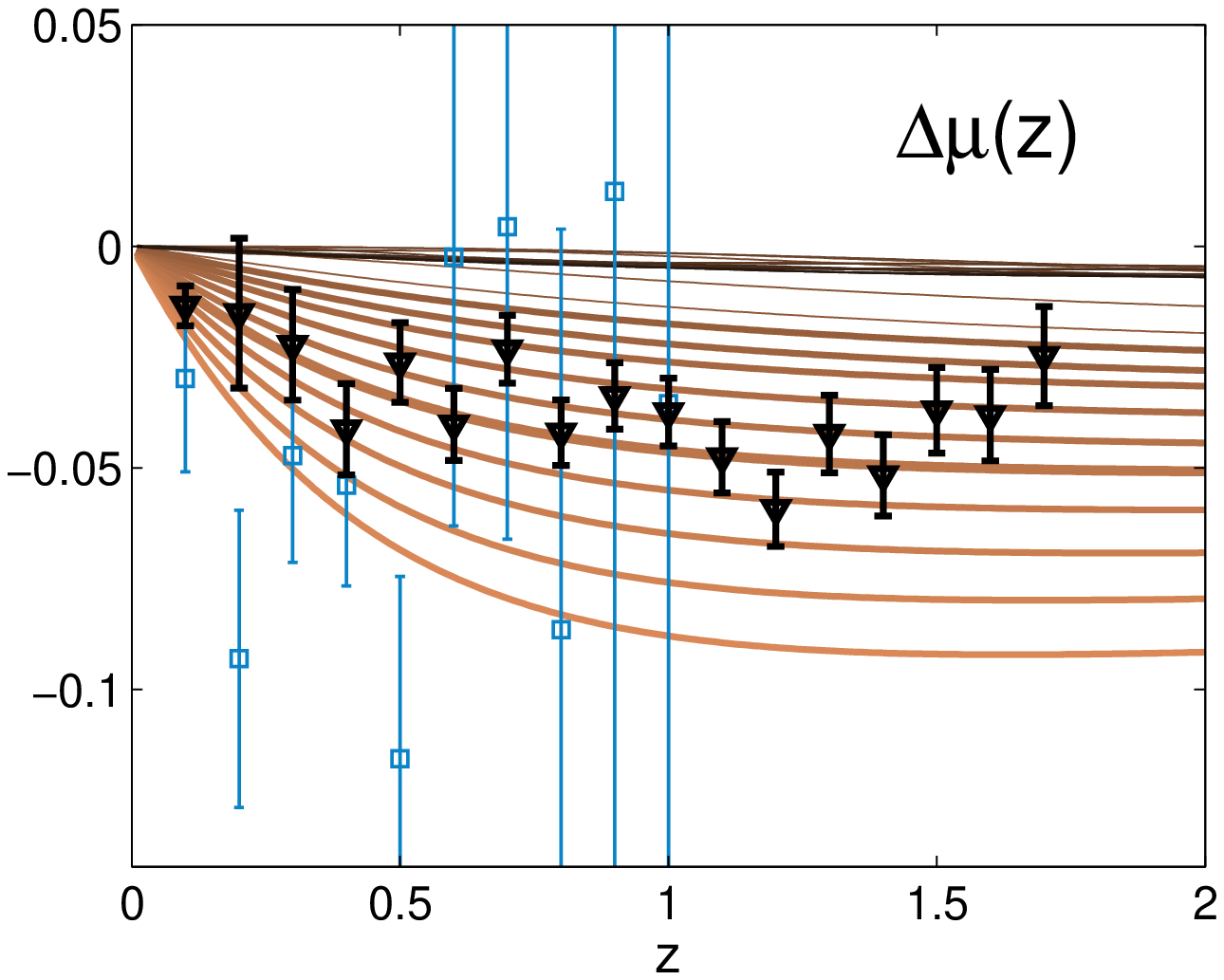} \\ [0.0cm]
 \end{array}$
\end{flushleft}
\caption{{\bf BBN compatible scaling dark energy models:} allowed observables for the polynomial $w(z)$ for
$z_t$ from $4.02$ (light brown) to $10.02$ (dark brown) {(\bf top panel)} and double exponential potential with
$-30 \leq \mu \leq 1$ {(\bf bottom)} showing the small deviations from the $\Lambda$ predictions. The bottom
light brown curves are from $\mu = 1$ (for which $w(0) = -0.8$) to $\mu = 0, w(0) = -1$ and the dark brown
curves are for $\mu = -30, -25, -20, -15, -10, -5, -4, -3, -2, -1, -0.8, -0.6, -0.4, -0.2$. All models have
$|dw/dz(0)| < 0.2$. In the model with $w(0) < -0.9$  $H_{\rm DE}(z)$ deviates from $H_{\Lambda}(z)$ by at most $2.7\%$ (marked
by the horizontal line in the bottom middle panel) and implies a deviation in distance modulus of less than 0.05
mag. The error bars in the right panels correspond to the Stage-III (large boxed errors) and Stage-IV (small
triangular errors) supernova surveys respectively. These are produced for the bottom curve in each case except
for the Stage-IV (SNAP-like) errors in the double exponential case which correspond to the $w(0)=-0.9$ model
(thicker line asymptoting to $\sim 0.04$ mag). Note that the ratio $H(z)/H_{\Lambda}$ for the double exponential
potential does not converge to $1.023$ since the matter-dominated value of $\Omega_{\rm DE}(z)$ is constrained
to be $3/4$ of the radiation-dominated value. Hence the Hubble rate is forced to $\sim 1 + (3/8)\epsilon \sim
1.017$. } \label{group1}
\end{figure*}

In the case of scaling dark energy models these constraints at early times are preserved by the subsequent
evolution of the cosmos, implying that the limit $\Omega_{\rm{DE}}<  \epsilon = 0.045$ holds for all redshifts
greater than $z_t$, the redshift at which the equation of state stops scaling and becomes negative.  We find new
implications for the magnitude of allowed deviations of these scaling models from $\Lambda$. To motivate our
results consider first a toy, step-function model for $w(z)$ in which $w$ is constant for $z < z_t$ and $w=0$
for $z \geq z_t$. Requiring $\Omega_{\rm DE}(z_t) < \epsilon = 0.045$ (the BBN constraint preserved by the
scaling field) with $w \geq -1$ and $\Omega_{\rm DE} =  0.7$ today implies that $z_t > 2.6$, an unexpectedly
large number given that most cosmic probes in the next decade will be limited to $z < 2$. To achieve $z_t = 1$
one requires instead  $w = -1.88$, a value disfavoured by current observations \cite{wmap3}.
Figure~(\ref{group1}) shows  that the early dark energy constraints on the scaling models we consider force the
derivative of $w(z)$ to be very small: $|w'(0)| < 0.2$ for all our models, significantly smaller than will be
detectable in the next decade \cite{detf}. If $w(0) < -0.9$ the deviation of the Hubble rate from $\Lambda$ is
less than $3\%$ and the deviation of the distance modulus is less than 0.04 mag. In addition, we find the
surprising result that the standard CPL parametrisation \cite{cp} cannot be used to describe scaling minimally
coupled scalar fields (which automatically have $w \geq -1$) and simultaneously match the nucleosynthesis bound.
Throughout the paper we assume a flat universe with $\Omega_m = 0.3$ today and all results are valid for
redshifts less than matter-radiation equality. We start by describing the constraints on general models in
Section \ref{gen_result}, we give the specific scaling models we consider in Section \ref{scaling_models} and
discuss the performance of standard parameterisations in describing these scaling fields in Section
\ref{cpl_section}. We conclude in Section \ref{conclude}.

\section{General Results\label{gen_result}}
One can derive model-independent constraints on cosmological observables (we consider the Hubble rate, $H(z)$,
and distance modulus, $\mu(z)$) for scaling models. The BBN constraint implies  that $\Omega_{\rm DE}(z\geq
z_{\rm t}) < \epsilon = 0.045$ since $\Omega_{\rm DE}(z\geq z_t)$ is constant in perfectly scaling models.
It is standard to write $\rho_{\rm DE}(z) = H_0^2\Omega_{\rm DE}(z=0) f(z)$ where
\begin{equation}
f(z) \equiv \exp \left[3\int_0^z \frac{1+w(z')}{1+z'} dz' \right]
\end{equation}
determines the redshift dependence of the dark energy density.
The evolution of $\Omega_{\rm DE}$ as a function of redshift is illustrated for various models in Figure~(\ref{ode}).
From the above discussion, plus the scaling requirement that $w(z \geq z_t) = 0$, one can show that
\begin{equation}
f(z\geq z_t) = \frac{\epsilon (1+z)^3}{r(1-\epsilon)} = \frac{0.047}{r}(1+z)^3
\end{equation}
where $r = \Omega_{\rm DE}/\Omega_m$ is evaluated today.

Using this result implies the following general but stringent constraint:
\begin{eqnarray}
\frac{H_{\rm DE}}{H_{\Lambda}}(z \geq z_t) &=& \sqrt{\frac{(1+z_t)^3}{(1-\epsilon)((1+z_t)^3 + r)}} \nonumber \\
&\leq& \sqrt{\frac{1}{1-\epsilon}}=1.023\,. \label{hlimit}
\end{eqnarray}
where the last equality arises from imposing $\epsilon = 0.045$ and the upper bound from setting $r=0$ as $z_t
\rightarrow \infty$. This limit can clearly be seen in the top middle panel of Fig.~(\ref{group1}). This robust
result implies that detecting deviations from $\Lambda$ at high redshift will be difficult using Hubble rate
measurements alone since they are bound to be less than $2.3\%$. This does not, however, strongly constrain the
behaviour of $H_{\rm DE}/H_{\Lambda}$ for $z < z_t$, once the field has left the scaling regime. We will show in
two classes of models that its maximum value is less than $5\%$ and occurs around $z \simeq 1$. The latter
result is good news for Baryon Acoustic Oscillation (BAO) surveys such as WiggleZ, BOSS and WFMOS \cite{wfmos}
which will probe this range of redshifts.

Similarly one can also place robust bounds on the deviation of the distance modulus, $\mu(z)$, from the
$\Lambda$ prediction.  The quantity $\Delta \mu \equiv \mu_{\rm DE}(z) - \mu_{\Lambda}(z)$ is given by:
\begin{equation}
\Delta \mu = 5 \log_{10}\left(\frac{d_{L,\rm DE}(z)}{d_{L,\Lambda}(z)}\right)
\end{equation}
where $d_L(z)$ is the corresponding model luminosity distance. If we assume that there exists a number $\alpha$
such that $H(z)/H_{\Lambda}(z) \leq 1 + \alpha^2$ for all $z$ (for example in the constraint on $H$ discussed
above $\alpha^2 \simeq 2.3 \%$) then $d_{L,\rm DE}(z)/d_{L,\Lambda}(z) \geq (1 + \alpha^2)^{-1}$ and hence
$\Delta\mu(z)$ obeys the inequality
\begin{equation}
0 \geq \Delta \mu(z) \geq - 5\log_{10}(1 + \alpha^2)
\end{equation}
A bound of $\alpha^2 = 0.025$ gives $|\Delta \mu(z)| \leq 0.054$ mag - see the top row of Fig.~(\ref{group1}).
This is a conservative upper bound since we have shown that for $z\geq z_t$,  $\alpha^2 \leq \epsilon/2$. For
the oscillating double exponential $w(z)$ models (with $-30\leq \mu < 0$) considered later, one has $\alpha^2 <
0.015$ which yields the constraint $|\Delta \mu(z)| \leq 0.032$ mag. These general results do not constrain
$H(z)$ for $z < z_t$, which instead requires a specific model for $w(z)$. We now consider two classes of models
which describe a wide range of scalar field dynamics.


\section{\label{scaling_models}Specific Forms of the Scaling Potential}

\subsection{Polynomial w(z) parametrisation}

 First we consider a quadratic parametrisation of the dark energy equation of state, $w(z)$ \cite{Weller:2002a}:
\begin{equation}
w(z) =
\begin{cases}
w_0 + w_1 z + w_2 z^2 & \text{for $z < z_t$} \\
0 & \text{for $z \geq z_t $}
\end{cases}
\end{equation}
We apply the constraint $w(z) \geq -1$ since we want to describe minimally coupled scalar fields with canonical
kinetic terms. The linear case with $w_2=0$ requires $z_t \simeq 6.2$ to match BBN (for $\epsilon  = 0.045$) if
$w_0 = -1$ and if we allow $w_0$ to be free the BBN constraint implies the correlation $w_1 \simeq -0.4 w_0 -
0.2$ for the interesting region $-1 < w_0 < -0.8$. The other case we consider is $w_0 = -1, w_2 \neq 0$.
Continuity at $z=z_t$ then implies \begin{equation}w_2 = \frac{1}{z_t^2} - \frac{w_1}{z_t}.\end{equation} The
BBN constraint provides $w_1$ in terms of $z_t$. The resulting family of curves and observables are shown in the
top row of Fig.~(\ref{group1}). We note that for $z \leq 1$ the BBN constraint is so strong that all models have
$w(z) < -0.8$, as shown in Fig.~(\ref{group1}), and the largest deviation of $H(z)$ from the $\Lambda$
comparison model is about $2.7\%$ with the largest deviation in the distance modulus only about 0.03 magnitudes
(occurring at $z = 2$). This shows that if $w \simeq -1$ today we cannot expect significant deviations from
$\Lambda$ at any redshift and only Stage-IV experiments \cite{detf} are likely to detect dark energy dynamics
with any real significance. However while current data favour $w(0) \simeq -1$ they are consistent with larger
values. To study how this zero-point affects our results we now consider simulations of a scalar field with a
popular family of scalar field potentials, $V(\phi)$, which also allows us to examine the impact of oscillations
in $w(z)$.

\subsection{Double Exponential Potential}
   A single exponential potential is well known to give early scaling \cite{copeland_tsujikawa} but cannot
   also lead to late-time acceleration. One well-studied way to combine the two is via the double exponential
   potential \cite{Barreiro:1999zs}:
\begin{equation}
V(\phi)=M_1^4 e^{-\lambda\kappa\phi}+M_2^4 e^{-\mu\kappa\phi}, \label{dep}
\end{equation}
where late-time acceleration is induced if $\mu \leq \sqrt{2}.$ The dynamics of $\phi$ is sensitive to the sign
of $\mu$, in that negative values of $\mu$ yield a potential that oscillates around the minimum of the field.
This can clearly be seen in Fig.~\ref{group1}, for the potentials with $-30 \leq \mu \leq 0$ We assume that the
scalar field is in the scaling regime during radiation domination, and hence satisfying the BBN constraint
requires $\lambda \geq 2/\sqrt{\epsilon} \geq 9.43$ \cite{copeland_tsujikawa}. We choose $\lambda = 9.43$ to
maximise deviations from $\Lambda$. Unlike a perfectly scaling model of the form we assumed in the previous
section, $\Omega_{\phi}$ actually decreases in the transition to matter domination and we have
$\Omega_{\phi}<3/4\times \epsilon = 0.034$ during matter domination, leading to even more stringent results -
see Fig.~(\ref{ode}). The extra $3/4$ factor is specific to the double exponential potential and is responsible
for the reduction of the asymptotic values of $H_{\rm DE}(z)/H_{\Lambda {\rm CDM}}$, seen in the bottom middle
panel of Fig.~(\ref{group1}), relative to the predictions of Eq.\eqref{hlimit}. We choose $M_1=10^{-14} m_{\rm pl}$ and use
Planck units where $\kappa=1$.

We numerically solve the evolution equation \begin{equation}\ddot{\phi}+3H\dot{\phi}+V_{,\phi}=0\end{equation}
for $\phi$ with $V(\phi)$ given by Eq.\eqref{dep}, together with the radiation and matter fluids. These are all
coupled to gravity through the Friedmann equation
\begin{equation}H^2=\frac{\kappa^2}{3}\left[\frac{1}{2}\dot{\phi}^2+V(\phi)+\rho_r+\rho_m\right].\end{equation} For each
value of $\mu$ we find $M_2$ such that $\Omega_{\rm DE} = 0.7$ today which implies $M_2\sim10^{-31} m_{\rm pl}$
for $\mu \sim 1$. The resulting $w(z)$ curves for these models, together with the predicted observables ($H(z),
\Delta \mu(z)$) are shown in the bottom row of Fig.~(\ref{group1}). For negative $\mu$  - we study
$-30\leq\mu<0$ - the potential has a global minimum and for $z\lesssim 0.2$ the equation of state satisfies
$w_{\phi}\leq -0.98$. As a result all the negative $\mu$ models show tiny deviations from $\Lambda$: less than
$1.5\%$ for $H(z)$ and less than 0.015 mag for $\Delta \mu$ (see Fig.~(\ref{group1})). This will make detection
extremely difficult even with the Stage-IV dark energy experiments such as DUNE, JDEM, LSST and SKA \cite{detf}.

In contrast, positive values of $\mu$ ($0\leq\mu\leq1$) simply modify the slope of the potential. They can yield
values of $w_{\phi}$ significantly different from $-1$ today, e.g. for $\mu \sim 1$ one finds
$w_{\phi}(0)\sim-0.8$ which is consistent (at about the 2$\sigma$ level \cite{wmap3, longde}) with current
observations and which we therefore take as the upper bound for $\mu$. The bottom row of Fig.~(\ref{group1})
shows that the maximum allowed deviation for $H(z)$ from $H_{\Lambda}(z)$ in this case is about $5\%$, peaking
at $z \sim 1$ with a maximum value of $\Delta \mu \sim 0.9$ mag. Such a model will be detectable with Stage-III
supernova surveys (at the $99.97\%$ confidence level) and with the upcoming BAO experiments, since the maximum
deviation in $H(z)$ coincides with the redshift ranges in which they will operate, i.e. $z \sim 0.7$ to $1.1$.
However, for values more consistent with the current best-fits, $w_{\phi}(0) < -0.9$ one finds much smaller
deviations of $2.7\%$ and 0.045 mag respectively for $H(z)$ and $\Delta \mu$ which again will require Stage-IV
experiments for conclusive detection as can be seen in the right-hand panels of Fig.~(\ref{group1}). Similar
results will apply to other modifications of the exponential potential, e.g. \cite{sahni}.

\section{Performance of standard parametrisations \label{cpl_section}}
  The most widely used parametrisation for dark energy, the Chevallier-Polarski-Linder (CPL) parametrisation \cite{cp},
   \begin{equation} w(z)=w_0+w_a\frac{z}{1+z},\end{equation} is also the basis for the DETF figure of merit \cite{detf}.
    Surprisingly this parametrisation fails dramatically to meet the BBN constraint if we demand $w(z) \geq -1$,
    $\Omega_{\rm DE} \sim 0.7$ today and $w(z\geq z_t)=0$ as before (for some $z_t$). This is clearly visible in
    the top curve of Fig.~(\ref{ode}), which shows the lowest attainable value of $\Omega_{\rm DE}$ with $w_0 = -1$.
    In retrospect this is understandable since to reach the scaling value $w=0$ for some $z_t$ requires $w_a > -w_0$.
    However in this case $w(z)$ doesn't spend enough time at sufficiently negative values to force $\Omega_{\rm DE}(z)$
    down to the BBN value.  The least phantom value of $w_0$ that is able to satisfy the BBN constraint is $w_0 = -1.3$.
    In contrast the logarithmic expansion $w(z) = w_0 + w_1 \ln(1+z)$ is able to match the BBN constraint with
    $w(z) \geq -1$, but only for $z_t > 12.4 $.


\section{Conclusions \label{conclude}}

Scaling field models are arguably the best-motivated alternatives to the cosmological constant. We have shown
that the constraints on the energy density of the scalar field at the time of Big Bang Nucleosynthesis and
decoupling strongly limit their allowed dynamics. If $w$ today is not close to the maximum value allowed by
current data then detection of dynamics will likely have to wait a decade for the Stage-IV DETF experiments. Of
course, these strong conclusions are only true for scaling models and if one allows exotic phantom behaviour ($w
< -1$) the conclusion is much more rosy. We discussed two specific families of these scaling models -  imposing
the BBN and CMB constraints on general scaling models is left for future work.

One might ask how the constraints discussed here will affect growth of structure. Since the growth in
minimally coupled scalar fields smoothly approaches that of $\Lambda$CDM when $w(z) \rightarrow -1$
(see e.g. Fig. 2 in \cite{MCBW}) our results show
that the growth in models allowed by BBN and CMB constraints will be close to those in $\Lambda$CDM at $z < 1$,
making detection difficult. To what
extent there are deviations at higher redshift, and whether they will be detectable by future cluster or weak lensing
surveys is again left as a subject for future study.

Finally we have shown that the
standard CPL parametrisation, $w(z) = w_0 + w_a z/(1+z)$, fails dramatically to match the BBN constraint when
describing scaling fields which satisfy $w \geq -1$. This is particularly important given that the CPL
parametrisation is the basis of the DETF figure of merit \cite{detf} which is now the {\em de facto} standard
for the optimisation of future cosmological surveys, e.g. \cite{parkinson}.  A concern therefore is that
optimisations may be unwittingly biased {\em away} from scaling dark energy models. More work in this area is
clearly needed to assess the implications for cosmological survey design, but it is clear that the current
non-detection of dark energy dynamics should neither come as a surprise, nor should it discourage us from the
hunt.

\section{Acknowledgements} We thank Chris Blake, Pier Stefano Corasaniti, Martin Kunz, Andrew Liddle,
David Parkinson, Martin Sahl\'en and Jussi Valiviita for discussions and the JCAP referee for useful suggestions.
This work was supported by the NRF, KAT, UCT, NASSP and FCT(Portugal).

\bibliographystyle{apsrev1}
\bibliography{bbn_combined}

\begin{thebibliography}{21}
\expandafter\ifx\csname natexlab\endcsname\relax\def\natexlab#1{#1}\fi
\expandafter\ifx\csname bibnamefont\endcsname\relax
  \def\bibnamefont#1{#1}\fi
\expandafter\ifx\csname bibfnamefont\endcsname\relax
  \def\bibfnamefont#1{#1}\fi
\expandafter\ifx\csname citenamefont\endcsname\relax
  \def\citenamefont#1{#1}\fi
\expandafter\ifx\csname url\endcsname\relax
  \def\url#1{\texttt{#1}}\fi
\expandafter\ifx\csname urlprefix\endcsname\relax\def\urlprefix{URL }\fi
\providecommand{\bibinfo}[2]{#2}
\providecommand{\eprint}[2][]{\url{#2}}

\bibitem[{\citenamefont{{Spergel} et~al.}(2007)\citenamefont{{Spergel}, {Bean},
  {Dor{\'e}}, {Nolta}, {Bennett}, {Dunkley}, {Hinshaw}, {Jarosik}, {Komatsu},
  {Page} et~al.}}]{wmap3}
\bibinfo{author}{\bibfnamefont{D.~N.} \bibnamefont{{Spergel}}},
  \bibinfo{author}{\bibfnamefont{R.}~\bibnamefont{{Bean}}},
  \bibinfo{author}{\bibfnamefont{O.}~\bibnamefont{{Dor{\'e}}}},
  \bibnamefont{et~al.}, \bibinfo{journal}{\apjs}
  \textbf{\bibinfo{volume}{170}}, \bibinfo{pages}{377} (\bibinfo{year}{2007}),
  \bibinfo{note}{{}P. Astier, J. Guy, N. Regnault, et. al., AAP {\bf 447}, 31
  (2006), M. Tegmark, D. Eisenstein, M. Strauss et. al., Phys. Rev. D {\bf 74},
  12 (2006), W. Wood-Vasey, G. Miknaitis, C. Stubbs et. al., (2007)
  astro-ph/0701041}.

\bibitem[{\citenamefont{{Armendariz-Picon}
  et~al.}(2000)\citenamefont{{Armendariz-Picon}, {Mukhanov}, and
  {Steinhardt}}}]{kessence1}
\bibinfo{author}{\bibfnamefont{C.}~\bibnamefont{{Armendariz-Picon}}},
  \bibinfo{author}{\bibfnamefont{V.}~\bibnamefont{{Mukhanov}}},
  \bibnamefont{and} \bibinfo{author}{\bibfnamefont{P.~J.}
  \bibnamefont{{Steinhardt}}}, \bibinfo{journal}{\prl}
  \textbf{\bibinfo{volume}{85}}, \bibinfo{pages}{4438} (\bibinfo{year}{2000}),
  \bibinfo{note}{{}C. Bonvin, C. Caprini and R. Durrer, Phys. Rev. Lett. {\bf
  97}, 8 (2006)}.

\bibitem[{\citenamefont{{Abramo} and {Pinto-Neto}}(2006)}]{k-essence_sound}
\bibinfo{author}{\bibfnamefont{L.~R.} \bibnamefont{{Abramo}}} \bibnamefont{and}
  \bibinfo{author}{\bibfnamefont{N.}~\bibnamefont{{Pinto-Neto}}},
  \bibinfo{journal}{\prd} \textbf{\bibinfo{volume}{73}},
  \bibinfo{pages}{063522} (\bibinfo{year}{2006}),
  \eprint{arXiv:astro-ph/0511562}.

\bibitem[{\citenamefont{Dvali et~al.}(2000)\citenamefont{Dvali, Gabadadze, and
  Porrati}}]{dgp1}
\bibinfo{author}{\bibfnamefont{G.~R.} \bibnamefont{Dvali}},
  \bibinfo{author}{\bibfnamefont{G.}~\bibnamefont{Gabadadze}},
  \bibnamefont{and} \bibinfo{author}{\bibfnamefont{M.}~\bibnamefont{Porrati}},
  \bibinfo{journal}{Phys. Lett.} \textbf{\bibinfo{volume}{B484}},
  \bibinfo{pages}{112} (\bibinfo{year}{2000}), \bibinfo{note}{{}C. Deffayet, G.
  Gabadadze and A. Iglesias, JCAP {\bf 8}, 12 (2006), C. Charmousis, R.
  Gregory, N. Kaloper et. al., JHEP {\bf 10}, 66 (2006)}.

\bibitem[{\citenamefont{{Koyama}}(2007)}]{dgp_ghosts}
\bibinfo{author}{\bibfnamefont{K.}~\bibnamefont{{Koyama}}},
  \bibinfo{journal}{Classical and Quantum Gravity}
  \textbf{\bibinfo{volume}{24}}, \bibinfo{pages}{231} (\bibinfo{year}{2007}),
  \eprint{arXiv:0709.2399}.

\bibitem[{\citenamefont{{Song} et~al.}(2007)\citenamefont{{Song}, {Peiris}, and
  {Hu}}}]{SPH}
\bibinfo{author}{\bibfnamefont{Y.-S.} \bibnamefont{{Song}}},
  \bibinfo{author}{\bibfnamefont{H.}~\bibnamefont{{Peiris}}}, \bibnamefont{and}
  \bibinfo{author}{\bibfnamefont{W.}~\bibnamefont{{Hu}}}
  (\bibinfo{year}{2007}), \bibinfo{note}{arXiv:0706.2399, T. Faulkner, M.
  Tegmark, E. Bunn et. al., (2006) astro-ph/0612569}.

\bibitem[{\citenamefont{{Sandvik} et~al.}(2004)\citenamefont{{Sandvik},
  {Tegmark}, {Zaldarriaga}, and {Waga}}}]{sand}
\bibinfo{author}{\bibfnamefont{H.~B.} \bibnamefont{{Sandvik}}},
  \bibinfo{author}{\bibfnamefont{M.}~\bibnamefont{{Tegmark}}},
  \bibinfo{author}{\bibfnamefont{M.}~\bibnamefont{{Zaldarriaga}}},
  \bibnamefont{et~al.}, \bibinfo{journal}{\prd} \textbf{\bibinfo{volume}{69}},
  \bibinfo{pages}{123524} (\bibinfo{year}{2004}).

\bibitem[{\citenamefont{{Ratra} and {Peebles}}(1988)}]{ratra}
\bibinfo{author}{\bibfnamefont{B.}~\bibnamefont{{Ratra}}} \bibnamefont{and}
  \bibinfo{author}{\bibfnamefont{P.~J.~E.} \bibnamefont{{Peebles}}},
  \bibinfo{journal}{\prd} \textbf{\bibinfo{volume}{37}}, \bibinfo{pages}{3406}
  (\bibinfo{year}{1988}), \bibinfo{note}{{}C. Wetterich, Nucl. Phys. B {\bf
  302}, 668 (1988), E. Copeland, A. Liddle and D. Wands, Phys. Rev. D {\bf 57},
  4686 (1998), P.~G. Ferreira and M. Joyce, Phys. Rev. Lett. {\bf 79}, 4740
  (1997)}.

\bibitem[{\citenamefont{{Copeland} et~al.}(2006)\citenamefont{{Copeland},
  {Sami}, and {Tsujikawa}}}]{copeland_tsujikawa}
\bibinfo{author}{\bibfnamefont{E.~J.} \bibnamefont{{Copeland}}},
  \bibinfo{author}{\bibfnamefont{M.}~\bibnamefont{{Sami}}}, \bibnamefont{and}
  \bibinfo{author}{\bibfnamefont{S.}~\bibnamefont{{Tsujikawa}}},
  \bibinfo{journal}{Int. J. Mod. Phys. D} \textbf{\bibinfo{volume}{15}}
  (\bibinfo{year}{2006}).

\bibitem[{\citenamefont{{Ferreira} and {Joyce}}(1998)}]{ferreira97}
\bibinfo{author}{\bibfnamefont{P.~G.} \bibnamefont{{Ferreira}}}
  \bibnamefont{and} \bibinfo{author}{\bibfnamefont{M.}~\bibnamefont{{Joyce}}},
  \bibinfo{journal}{\prd} \textbf{\bibinfo{volume}{58}},
  \bibinfo{pages}{023503} (\bibinfo{year}{1998}).

\bibitem[{\citenamefont{{Bean} et~al.}(2001)\citenamefont{{Bean}, {Hansen}, and
  {Melchiorri}}}]{Bean:2001wt}
\bibinfo{author}{\bibfnamefont{R.}~\bibnamefont{{Bean}}},
  \bibinfo{author}{\bibfnamefont{S.~H.} \bibnamefont{{Hansen}}},
  \bibnamefont{and}
  \bibinfo{author}{\bibfnamefont{A.}~\bibnamefont{{Melchiorri}}},
  \bibinfo{journal}{\prd} \textbf{\bibinfo{volume}{64}},
  \bibinfo{pages}{103508} (\bibinfo{year}{2001}).

\bibitem[{\citenamefont{{Doran} et~al.}(2005)\citenamefont{{Doran}, {Karwan},
  and {Wetterich}}}]{doran_cmb1}
\bibinfo{author}{\bibfnamefont{M.}~\bibnamefont{{Doran}}},
  \bibinfo{author}{\bibfnamefont{K.}~\bibnamefont{{Karwan}}}, \bibnamefont{and}
  \bibinfo{author}{\bibfnamefont{C.}~\bibnamefont{{Wetterich}}},
  \bibinfo{journal}{\jcap} \textbf{\bibinfo{volume}{11}}, \bibinfo{pages}{7}
  (\bibinfo{year}{2005}), \bibinfo{note}{{} M. Doran and G. Robbers, JCAP {\bf
  6}, 26 (2006)}.

\bibitem[{\citenamefont{Corasaniti et~al.}(2004)\citenamefont{Corasaniti, Kunz,
  Parkinson, Copeland, and Bassett}}]{longde}
\bibinfo{author}{\bibfnamefont{P.~S.} \bibnamefont{Corasaniti}},
  \bibinfo{author}{\bibfnamefont{M.}~\bibnamefont{Kunz}},
  \bibinfo{author}{\bibfnamefont{D.}~\bibnamefont{Parkinson}},
  \bibnamefont{et~al.}, \bibinfo{journal}{\prd} \textbf{\bibinfo{volume}{70}},
  \bibinfo{pages}{083006} (\bibinfo{year}{2004}).

\bibitem[{\citenamefont{{Chevallier} and {Polarski}}(2001)}]{cp}
\bibinfo{author}{\bibfnamefont{M.}~\bibnamefont{{Chevallier}}}
  \bibnamefont{and}
  \bibinfo{author}{\bibfnamefont{D.}~\bibnamefont{{Polarski}}},
  \bibinfo{journal}{Int. J. Mod. Phys. D} \textbf{\bibinfo{volume}{10}},
  \bibinfo{pages}{213} (\bibinfo{year}{2001}), \bibinfo{note}{{}E. Linder,
  Phys. Rev. Lett. {\bf 20}, 9 (2003)}.

\bibitem[{\citenamefont{Albrecht et~al.}(2006)\citenamefont{Albrecht,
  Bernstein, Cahn, Freedman, Hewitt, Hu, HUTH, Kamionkowski, Kolb, Knox
  et~al.}}]{detf}
\bibinfo{author}{\bibfnamefont{A.}~\bibnamefont{Albrecht}},
  \bibinfo{author}{\bibfnamefont{G.}~\bibnamefont{Bernstein}},
  \bibinfo{author}{\bibfnamefont{R.}~\bibnamefont{Cahn}}, \bibnamefont{et~al.},
  \bibinfo{journal}{astro-ph/0609591}  (\bibinfo{year}{2006}).

\bibitem[{\citenamefont{Bassett et~al.}(2005)\citenamefont{Bassett, Nichol,
  Eisenstein, and Team}}]{wfmos}
\bibinfo{author}{\bibfnamefont{B.~A.} \bibnamefont{Bassett}},
  \bibinfo{author}{\bibfnamefont{R.~C.} \bibnamefont{Nichol}},
  \bibinfo{author}{\bibfnamefont{D.~J.} \bibnamefont{Eisenstein}},
  \bibnamefont{et~al.}, \bibinfo{journal}{A\&G} \textbf{\bibinfo{volume}{46}},
  \bibinfo{pages}{5.26} (\bibinfo{year}{2005}), \bibinfo{note}{~K. Glazebrook
  et al., astro-ph/0701876 (2007)}.

\bibitem[{\citenamefont{{Weller} and {Albrecht}}(2002)}]{Weller:2002a}
\bibinfo{author}{\bibfnamefont{J.}~\bibnamefont{{Weller}}} \bibnamefont{and}
  \bibinfo{author}{\bibfnamefont{A.}~\bibnamefont{{Albrecht}}},
  \bibinfo{journal}{\prd} \textbf{\bibinfo{volume}{65}},
  \bibinfo{pages}{103512} (\bibinfo{year}{2002}).

\bibitem[{\citenamefont{{Barreiro} et~al.}(2000)\citenamefont{{Barreiro},
  {Copeland}, and {Nunes}}}]{Barreiro:1999zs}
\bibinfo{author}{\bibfnamefont{T.}~\bibnamefont{{Barreiro}}},
  \bibinfo{author}{\bibfnamefont{E.~J.} \bibnamefont{{Copeland}}},
  \bibnamefont{and} \bibinfo{author}{\bibfnamefont{N.~J.}
  \bibnamefont{{Nunes}}}, \bibinfo{journal}{\prd}
  \textbf{\bibinfo{volume}{61}}, \bibinfo{pages}{127301}
  (\bibinfo{year}{2000}).

\bibitem[{\citenamefont{{Sahni} and {Wang}}(2000)}]{sahni}
\bibinfo{author}{\bibfnamefont{V.}~\bibnamefont{{Sahni}}} \bibnamefont{and}
  \bibinfo{author}{\bibfnamefont{L.}~\bibnamefont{{Wang}}},
  \bibinfo{journal}{\prd} \textbf{\bibinfo{volume}{62}},
  \bibinfo{pages}{103517} (\bibinfo{year}{2000}).

\bibitem[{\citenamefont{{Ma} et~al.}(1999)\citenamefont{{Ma}, {Caldwell},
  {Bode}, and {Wang}}}]{MCBW}
\bibinfo{author}{\bibfnamefont{C.-P.} \bibnamefont{{Ma}}},
  \bibinfo{author}{\bibfnamefont{R.~R.} \bibnamefont{{Caldwell}}},
  \bibinfo{author}{\bibfnamefont{P.}~\bibnamefont{{Bode}}},
  \bibnamefont{et~al.}, \bibinfo{journal}{\apjl}
  \textbf{\bibinfo{volume}{521}}, \bibinfo{pages}{L1} (\bibinfo{year}{1999}),
  \eprint{arXiv:astro-ph/9906174}.

\bibitem[{\citenamefont{{Parkinson} et~al.}(2007)\citenamefont{{Parkinson},
  {Blake}, {Kunz}, {Bassett}, {Nichol}, and {Glazebrook}}}]{parkinson}
\bibinfo{author}{\bibfnamefont{D.}~\bibnamefont{{Parkinson}}},
  \bibinfo{author}{\bibfnamefont{C.}~\bibnamefont{{Blake}}},
  \bibinfo{author}{\bibfnamefont{M.}~\bibnamefont{{Kunz}}},
  \bibnamefont{et~al.}, \bibinfo{journal}{\mnras}
  \textbf{\bibinfo{volume}{377}}, \bibinfo{pages}{185} (\bibinfo{year}{2007}).

\end{thebibliography}

\end{document}